\documentclass[fleqn,usenatbib]{mnras}

\usepackage{newtxtext,newtxmath}
\usepackage[T1]{fontenc}
\usepackage{multirow}

\DeclareRobustCommand{\VAN}[3]{#2}
\let\VANthebibliography\thebibliography
\def\thebibliography{\DeclareRobustCommand{\VAN}[3]{##3}\VANthebibliography}

\usepackage{graphicx}	% Including figure files
\usepackage{amsmath}	% Advanced maths commands
\usepackage{soul}
\usepackage{xcolor}
\usepackage{ulem}

\title[High-resolution imaging of Dyson Sphere Candidate G]{High-resolution imaging of the radio source associated with Project Hephaistos Dyson Sphere Candidate G}

\author[T. Ren, M. A. Garrett and A. P. V. Siemion]{
T. Ren,$^{1}$\thanks{E-mail: tongtian.ren@manchester.ac.uk (TR)}
M. A. Garrett,$^{1,2,6}$\thanks{E-mail: michael.garrett@manchester.ac.uk (MAG)},
and A. P. V. Siemion$^{1,3,4,5,6}$
\\
$^{1}$ Jodrell Bank Centre for Astrophysics, Department of Physics and Astronomy, School of Natural Sciences, University of Manchester, Oxford Road,\\ Manchester M13 9PL, UK\\
$^{2}$ Leiden Observatory, Leiden University, PO Box 9513, NL-2300 RA Leiden, the Netherlands\\
$^{3}$ Breakthrough Listen, Astrophysics sub-department, Department of Physics, University of Oxford, Denys Wilkinson Building, Keble Road, Oxford OX1 3RH, UK\\
$^{4}$ Berkeley SETI Research Center, University of California, Berkeley, CA 94720, USA\\
$^{5}$ SETI Institute, 339 Bernardo Avenue, Suite 200, Mountain View, CA 94043, USA\\
$^{6}$ University of Malta, Institute of Space Sciences and Astronomy, Msida, MSD2080, Malta
}

\date{Accepted by MNRAS Letters on 2025 January 9; Received 2025 January 9; in original form 2024 December 17.}

\pubyear{2025}

\begin{document}
\label{firstpage}
\pagerange{\pageref{firstpage}--\pageref{lastpage}}
\maketitle

\begin{abstract}
We present high-resolution e-MERLIN and EVN (e-VLBI) observations of a radio source associated with Dyson Sphere candidate G, identified as part of Project Hephaistos. The radio source, VLASS J233532.86-000424.9, is resolved into three compact components and shows the typical characteristics of a radio-loud active galactic nucleus (AGN). In particular, the European VLBI Network (EVN) observations show that it has a brightness temperature in excess of $10^{8}$~K. No radio emission is detected at the position of the M-dwarf star. This result confirms our earlier hypothesis, that at least some of the Dyson Sphere candidates of project Hephaistos are contaminated by obscured, background AGN, lying close to the line of sight of otherwise normal galactic stars.  High-resolution radio observations of other Dyson Sphere candidates can be useful in distinguishing truly promising candidates from those contaminated by background sources. 

\end{abstract}


%%%%%%%%%%%%%%%%%%%%%%%%%%%%%%%%%%%%%%%%%%%%%%%%%%

%%%%%%%%%%%%%%%%% BODY OF PAPER %%%%%%%%%%%%%%%%%%

\section{Introduction}

Project Hephaistos recently published a list of Dyson Sphere candidates selected from 5 million sources with distances measured by {\it{Gaia}} to $< 300$~parsecs. Seven candidates were identified all of which are M-dwarf stars exhibiting an infrared excess in the WISE bands W3 and W4. These peculiar stellar Spectral Energy Distributions (SEDs) are consistent with simple Dyson Sphere models. \cite{2024MNRAS.531..695S} considered several natural explanations for the candidate SEDs, including warm debris disks which can also produce excess middle infrared (MIR) emission \citep{2016ApJS..225...15C}. However, the characteristics of these candidates challenge natural explanations: debris disks around M-dwarfs are extremely rare, and these candidates show higher temperatures and larger fractional luminosities than typical debris disks. While young stellar disks can have larger fractional luminosities, the seven candidates lack the variability typically associated with young stars. In addition, extreme debris disks (EDDs; \citet{2009ApJ...698.1989B}), which can have high fractional luminosities and higher temperatures \citep{2021ApJ...910...27M}, have never been observed around M-dwarfs. 

Extragalactic background contamination is a significant consideration for galactic SETI \citep{2020MNRAS.498.5720W,2023MNRAS.519.4581G}. Although \cite{2024MNRAS.531..695S} used the Improved Reprocessing of the IRAS Survey (IRIS; \citet{2005ApJS..157..302M}) maps at 100~$\mu\mathrm{m}$ to eliminate potential contamination, this method cannot reject all possible contaminants. \citet*{2024RNAAS...8..145R} identified radio emission with three of the seven Dyson Sphere candidates (A, B, and G) using data from various archival radio surveys. The radio sources were offset from the {\it{Gaia}} stellar positions by $\sim$~4.9, $\sim$~0.4 and $\sim$~5.0 arcseconds, respectively.

We found that the radio source associated with G was detected in multiple radio surveys (e.g. VLASS J233532.86-000424.9, NVSS J233532-000425, RACS J233532.8-000425). The steep spectral index of this source ($\alpha \sim$ -0.52) suggested synchrotron emission from a radio-loud active galactic nucleus (AGN) with extended jets. \citet{2024MNRAS.531..695S} also noted an offset of 5.59 arcseconds mostly in right ascension for candidate G as measured between the W1 and W3 bands. We therefore proposed \citep*{2024RNAAS...8..145R} that the MIR measurements associated with candidate G were probably contaminated by a background galaxy associated with VLASS J233532.86-000424.9. We further hypothesised that since the galaxy would have to be faint in the optical and near-infrared (NIR) but bright in the MIR, it was most likely to be contaminated by a hot, dust-obscured galaxy or "hot DOG" \citep{2015ApJ...804...27A}. 

In this way, the relatively low-resolution MIR measurements of candidate G could be significantly contaminated with a strong excess being observed in the MIR ({\it i.e.} the W3 and W4 bands of WISE). This excess would make this otherwise normal M-dwarf star look like an excellent DS candidate. We also suggested that the other candidates associated with radio sources (A \& B) might also be contaminated by background hot DOGs. We further reasoned in \citet*{2024RNAAS...8..145R} that the estimated surface density on the sky of hot DOGS \citep{2015ApJ...804...27A} might suggest all of the 7 Dyson Sphere candidates were simply M-dwarf stars contaminated by background hot DOGs. In a more careful analysis, \citet{2024arXiv240911447B} came to a similar conclusion, taking into account the positional uncertainties of the sources.

%but are fainter than VLASS J233532.86-000424.9. Candidates C, D, E, and F, which are not found to have radio counterparts, have their infrared emissions presumably contaminated by radio-quiet hot DOGs. 

In this paper, we present new high resolution images of the radio source associated with Dyson Sphere candidate G (VLASS J233532.86-000424.9, hereafter J2335-0004). In section \ref{sec2} we describe observations of J2235-0004 made independently by e-MERLIN at L-band and the European VLBI Network (EVN) at C-band. The data analysis is also summarised in this section. In section \ref{sec3} we present the results and these are further discussed in section \ref{sec4}. The conclusions are presented in section \ref{sec5}.

%We also analysed the VLA-SE and ALLWISE archival images to investigate the offset between the radio source and the infrared source. 

\section{Observations and Data Analysis}
\label{sec2}

\subsection{e-MERLIN data}

\begin{table*}
    \centering
    \caption{Details of the e-MERLIN and EVN observations. (a) Total on-source time excluding calibration overheads.}
    \begin{tabular*}{\textwidth}{@{\extracolsep{\fill}}lccccl@{}}
        \hline
        \textbf{Telescope} & \textbf{Project code} & \textbf{Obs. date} & \textbf{Time on source\textsuperscript{a}} & \textbf{Frequency} & \textbf{Calibrators}  \\
         &  &  & (h) & (GHz) & \\
        \hline
        \multirow{4}{*}{e-MERLIN} & \multirow{4}{*}{RR17003} & \multirow{4}{*}{June 7, 2024} & \multirow{4}{*}{9.5} & \multirow{4}{*}{1.25 -- 1.70} & J2335-0034 (phase calibrator) \\
        & & & & & J0319+4130 (bandpass calibrator) \\
        & & & & & J1331+3030 (flux calibrator) \\
        & & & & & J1407+2827 (additional calibrator) \\
        \hline
        \multirow{3}{*}{EVN} & \multirow{3}{*}{RSG19} & \multirow{3}{*}{June 19, 2024} & \multirow{3}{*}{2} & \multirow{3}{*}{4.81 -- 5.06} & J2253+1608 (fringe finder)\\
        & & & & & J2335-0034 (faint calibrator) \\
        & & & & & J2335-0131 (bright calibrator) \\
        \hline
    \end{tabular*}
    \label{tab:obs}
\end{table*}

We conducted observations of J2335-0004 using e-MERLIN at L-band (project code: RR17003). The observation took place on June 7, 2024, 05:00 to 14:34 UTC, using 6 antennas: Jodrell Bank Mark II, Pickmere, Darnhall, Knockin, Defford, and Cambridge. The observation was performed at L-band, covering a frequency range of 1.25-1.70~GHz,  centred at 1.48~GHz, with a total bandwidth of 512 MHz divided into 4 spectral windows. Each spectral window had 128 channels with a 1~MHz channel width.

The observations were conducted in phase-referencing mode. The target source J2335-0004 and the phase calibrator J2335-0034 were observed for about 6.5 hours with a cycle time of 10 minutes - a 7-minute scan on the target, followed by a 3-minute scan on the phase calibrator. 3C84 (J0319+4130) was used as a bandpass calibrator and 3C286 (J1331+3030) as the flux density calibrator. An additional calibrator, OQ208 (J1407+2827) was also observed. 

%\textbf{This paragraph was written according to the CASA pipeline log} 

The data were subsequently processed with the e-MERLIN \textsc{casa} Pipeline (v1.1.19) \citep{2021ascl.soft09006M}. The standard calibration steps included bandpass calibration using 3C84, flux density calibration with 3C286, and phase referencing with J2335-0034. The main limitation on the data was imposed by the poor uv-coverage for J2335-0004 with the source located close to the equator. The synthesised beam of the e-MERLIN data was 377~mas $\times$ 115~mas in position angle 23.8 $^{\circ}$. The final image from the pipeline is presented in Fig. \ref{fig:G}. The $1-\sigma$ noise level in the image is 0.275~mJy/beam. The triple structure of the source is clearly seen in the map.
%https://www.e-merlin.ac.uk/distribute/CY17/RR17003/RR17003_L_001_20240607/weblog/)

\subsection{EVN (e-VLBI) data}
We also conducted short EVN e-VLBI observations of J2335-0004 (project code: RSG19). The observations took place on June 19, 2024, 03:00 to 5:00 UTC, lasting $\sim$ 2 hours. 9 stations: Lovell (76m, UK), Westerbork (25m, Netherlands), Effelsberg (100m, Germany, reference antenna), Medicina (32m, Italy), Onsala (25m, Sweden), Torun (32m, Poland), Yebes (40m, Spain), Hartebeesthoek (26m, South Africa), and Irbene (32m, Latvia) participated in the observations. The observation was at C-band (centred at 4.94~GHz) with dual circular polarisation (LCP and RCP), and had a data rate of 2048 Mbps. The total bandwidth was 256 MHz (in both left and right-hand circular polarisation), including 8 spectral windows of 32 MHz each.

\begin{figure*}
    \centering
    \includegraphics[width=0.8\textwidth, keepaspectratio]{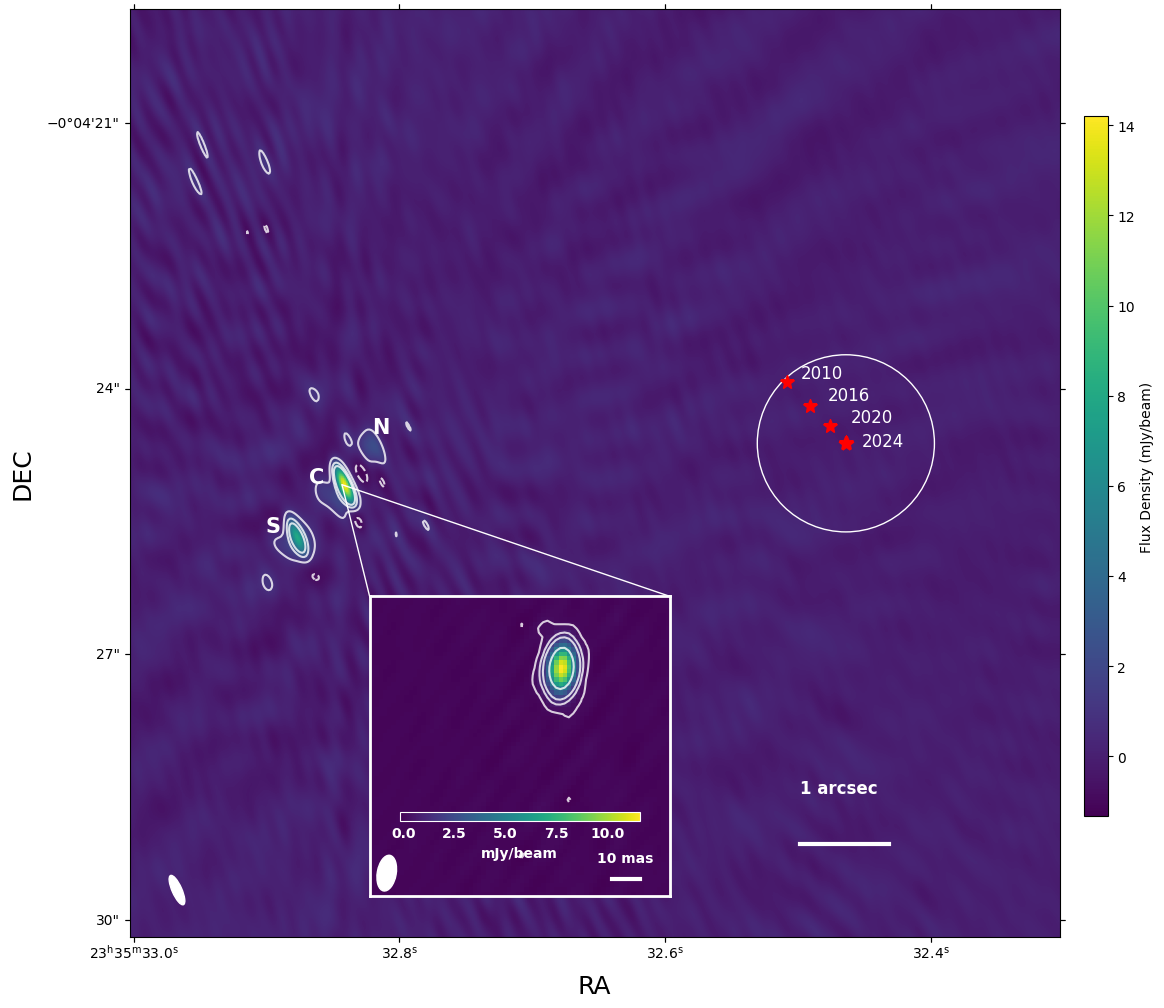}
    \caption{High-resolution e-MERLIN and EVN (e-VLBI) images of J2335-0004, with {\it{Gaia}} positions of candidate star G (red stars) marked within the white circle. For the e-MERLIN map, contours are drawn at $-1$, 1, 3, 5, and 30 times the $3\sigma$ noise level in the image (root mean square (RMS) = 0.825~mJy~beam$^{-1}$). The inset EVN image shows contours at $-1$, 1, 5, 10, 30, and 100 times its $3\sigma$ noise level (RMS = 0.165~mJy~beam$^{-1}$), with negative contours shown as dashed lines in both images.}
    \label{fig:G}
\end{figure*}

The observation was made in phase-reference mode using the same phase calibrator as employed by e-MERLIN: J2335-0034. We also used a brighter but more distant calibrator, J2335-0131 in the cycle. A cycle time of 7.5 minutes followed a pattern of 2 minutes on J2335-0131, 1.5 minutes on J2335-0034, and 3.5 minutes on the target. J2253+1608 (3C454.3) was used as a fringe finder. The data were correlated using the SFXC correlator at JIVE and processed using the standard EVN pipeline \citep*{2002astro.ph..5118R}. After the initial amplitude calibration and fringe-fitting, the corrections from the calibrators were applied to the target source. The data were then manually self-calibrated in three steps: an initial phase-only calibration with a 4-minute solution interval, followed by phase-only calibration with 30-second intervals, and a final amplitude and phase calibration using 4-minute intervals. The final image was produced using natural weighting. Figure \ref{fig:G} shows that a compact source was clearly detected by the EVN located close to the position of the peak in the e-MERLIN image.

Using the unaveraged visibility data, we also attempted to detect emission around the {\it{Gaia}} position of the M-dwarf star associated with candidate G - no detection was made. 

Table \ref{tab:obs} summarises the details of the e-MERLIN and EVN observations. Table \ref{tab:images} reports the main characteristics of the images presented here. 

\begin{table*}
\centering
\caption{The noise, peak brightness, SNR and beam size for different radio images.}
\label{tab:images}
\begin{tabular}{|l|c|c|c|c|}
\hline
\textbf{Observation} & \textbf{$1-\sigma$ r.m.s. noise} & \textbf{Peak brightness} & \textbf{SNR} & \textbf{Beam size} \\
 & \textbf{(mJy beam$^{-1}$)} & \textbf{(mJy beam$^{-1}$)} & & \textbf{[mas × mas ($\circ$)]} \\ \hline
e-MERLIN (1.5~GHz) & 0.275 & 14.198 & 51.7 & 376.6 $\times$ 115.0 (23.8) \\ 
EVN e-VLBI (5~GHz) & 0.055 & 11.599 & 209.3 & 12.7 $\times$ 6.6 (-10.0) \\ \hline
\end{tabular}
\end{table*}

\begin{figure*}
    \centering    \includegraphics[width=\textwidth, keepaspectratio]{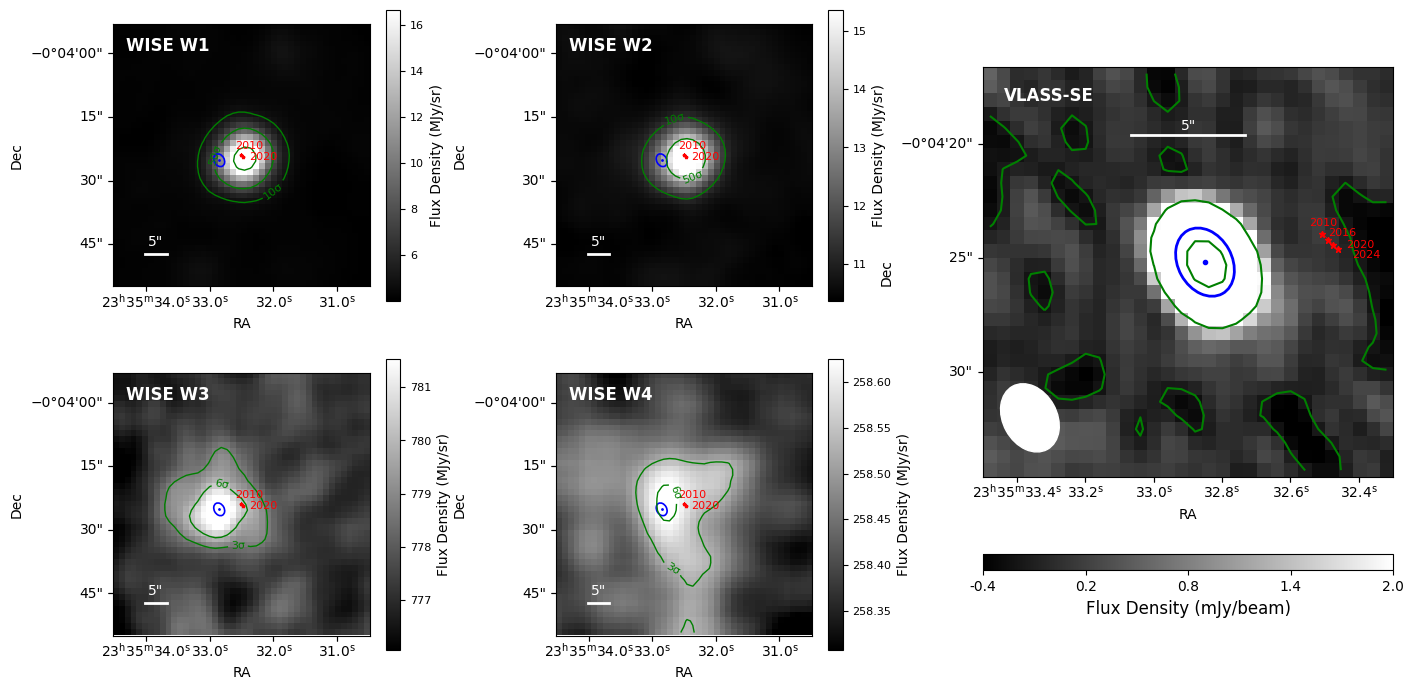}
    \caption{The WISE and VLASS-SE images of J2335-0004. The right panel is the VLASS-SE image. The full width at half maximum (FWHM) and centroid derived from the VLASS-SE image are blue ellipses and dots in all five panels. The four left panels are W1, W2, W3, and W4 band images of the source. The green contours represent $10\sigma$, $50\sigma$ and $100\sigma$ of the noise levels in WISE W1 and W2 images, $3\sigma$ and $6\sigma$ in W3 and W4 images, and $-1\sigma$, $10\sigma$ and $80\sigma$ in the VLASS-SE image. The candidate star's location is shown as the red stars in all the panels.}
    \label{fig:wise}
\end{figure*}

\section{Results}
\label{sec3}

\subsection{The e-MERLIN and e-VLBI Results}

In our high-resolution 1.5~GHz observations, e-MERLIN successfully resolved J2335-0004 into three components (see Figure \ref{fig:G}). We used the  \textsc{casa} \texttt{imfit} task to fit Gaussians to these data. A summary of the results is presented in Table \ref{tab:components}. Component C has the highest flux density ($\sim$ 14.06~mJy) and the highest brightness temperature of $\sim 1.79\times 10^5$~K. The other two components (S and N) exhibit lower flux densities ($\sim$ 9.94~mJy and $\sim$ 3.27~mJy) and lower brightness temperatures ($\sim 9.77\times 10^4$~K and $\sim 3.17\times 10^4$~K). The offsets of components S and N to C are $\sim 0.80$ and 0.54 arcseconds, respectively.

\begin{table*}
\centering
\caption{Properties of the Gaussian fits to the radio components in the e-MERLIN L-band and e-VLBI C-band images: flux density, brightness temperature, deconvolved angular size, and centroid positions.}
\label{tab:components}
\begin{tabular}{|l|c|c|c|c|}
\hline
\textbf{Component} & \textbf{Flux density} & \textbf{Brightness} & \textbf{Deconvolved size} & \textbf{Centroid position} \\
 & \textbf{(mJy)} & \textbf{temperature (K)} & \textbf{[mas × mas ($\circ$)]} & \textbf{RA, DEC} \\ \hline
C (1.5~GHz) & 14.06 $\pm$ 1.08 & (1.79 $\pm$ 0.18) $\times$ 10$^5$ & 341.8 $\times$ 124.7 (22.4) & 23:35:32.8414, -0:04:25.098\\
S (1.5~GHz) & 9.94 $\pm$ 0.83 & (9.77 $\pm$ 1.10) $\times$ 10$^4$ & 340.0 $\times$ 162.5 (19.8) & 23:35:32.8768, -0:04:25.690\\
N (1.5~GHz) & 3.27 $\pm$ 0.54 & (3.17 $\pm$ 0.71) $\times$ 10$^4$ & 309.2 $\times$ 181.0 (31.9) & 23:35:32.8207, -0:04:24.659\\ \hline
C (5~GHz) & 13.69 $\pm$ 0.12 &  (2.30 $\pm$ 3.18) $\times$ 10$^8$ & 4.46 $\times$ 0.67 (46.17) & 23:35:32.8406, -0:04:25.069\\ \hline
\end{tabular}
\end{table*}

At 5~GHz, the EVN resolved J2335-0004 into a single compact component located within 30~mas of the central component C detected by e-MERLIN. This offset is significantly smaller than the e-MERLIN restoring beam. The component's brightness temperature, derived from its EVN flux density of $13.69 \pm 0.12$~mJy at 5~GHz, is $ > 10^8$~K. Components S and N are not detected in the e-VLBI observations and are probably resolved by the much higher-resolution data.

Table \ref{tab:components} shows the properties of the components in two observations.

\subsection{The Offset between the radio and infrared sources}

We compared the positional offset between the radio source J2335-0004 and the WISE images of candidate G in the W1, W2, W3 and W4 bands. The Very Large Array Sky Survey Epoch 1 (VLASS-SE, \citet{2020PASP..132c5001L}) image of J2335-0004 is shown in Fig.~\ref{fig:wise}, along with the WISE multi-band images. We chose to make the comparison with the VLASS-SE since this image has a spatial resolution that is more similar to WISE. The 3.0~GHz VLASS-SE image shows an unsolved radio source with a total flux density of $27.4 \pm 0.3 \, \text{mJy}$. 

The VLASS-SE radio source is clearly offset to the east of the {\it{Gaia}} candidate star G by $\sim 5.64$ arcseconds. Also in Fig.~\ref{fig:wise}, we note that the {\it{Gaia}} position of candidate G agrees well with the centroid positions of the emission detected by WISE in the W1 and W2 bands. However, the peak of the mid-infrared emission in W3 and W4 is seen to move eastwards towards the VLA position of J2335-0004. For these two bands, the MIR and radio emission are essentially coincident with each other. We return to these results later in Section \ref{sec4}.

\subsection{The search for the X-ray counterparts}

We also searched for potential X-ray emission from J2335-0004 in the High Energy Astrophysics Science Archive Research Center (HEASARC) database\footnote{HEASARC: \url{https://heasarc.gsfc.nasa.gov/}}. J2335-0004 is included in the area of sky surveyed by Swift and ROSAT but no X-ray counterparts were found.

\section{Discussion}
\label{sec4}

The brightness temperature of the three e-MERLIN radio components of J2335-0004 are all in excess of ${1 \times 10^4}$~K (see Table \ref{tab:components}). In addition, the brightness temperature of the component C as measured by the EVN is in excess of ${1 \times 10^8}$~K. As noted by \citet{1991ApJ...378...65C}, radio emission from AGN activity typically has a brightness temperature of $> 10^{5}$~K. From the e-MERLIN 5~GHz and EVN 1.5~GHz flux density measurements for component C, we observe a flat spectrum core with a spectral index of $0.02 \pm 0.06$. This further strengthens the AGN interpretation of J2335-0004. The brightness temperature of the flat-spectrum core, together with the large-scale morphology of the source as seen by e-MERLIN, strongly argue that this radio source must be an AGN rather than radio emission from the M-dwarf star or some other galactic object.  

The fact that the centroid of the IR emission detected by WISE shifts across the W1-W4 bands is also telling. The centroids of the W1 and W2 band emission are consistent with the {\it{Gaia}} position of the M-dwarf Dyson Sphere candidate G but the positions of the emission in W3 and W4 are clearly offset from G and shift several arcseconds towards the position of J2335-0004. This suggests that while the M-dwarf star dominates the IR emission in the W1 and W2 bands, the background AGN (associated with radio source J2335-0004) dominates the MIR emission in bands W3 and W4. In other words, the background AGN is contaminating the W3 and W4 measurements that led \citet{2024MNRAS.531..695S} to identify this otherwise normal M-dwarf star as a potential Dyson Sphere candidate. 

The scenario presented here is consistent with our earlier claims that at least some (and possibly all) of the Project Hephaistos candidates are contaminated by background AGN. Since these candidates are all selected by significant deviations from a black-body stellar spectrum in the W3 and W4 bands, the background AGN must be faint in the W1 and W2 bands, but bright in W3 and W4. This observation reinforces our belief that this background AGN is a dust-obscured galaxy that is faint in the optical and NIR but bright in the MIR. Specifically, we identify these background, potentially contaminating AGN as “hot DOGs” \citep{2015ApJ...804...27A}. The fact that no X-ray emission is associated with J2335-0004 can be attributed to the AGN being Compton-thick, where the dense surrounding material absorbs the majority of X-ray radiation. This characteristic is consistent with the nature of “hot DOGs,” which are heavily obscured by dust and often exhibit weak or undetectable X-ray emission due to high column densities of intervening material ({\it e.g.}, the NuSTAR and {\it {XMM-Newton}} Observations for three hot DOGs in \citet{2014ApJ...794..102S}).

As shown in Figure \ref{fig:wise}, the M-dwarf star G dominates the emission in W1 and W2. While in the W3 and W4 bands, the brightest regions of the infrared are associated with the radio source. The M-dwarf star's location is located close to at an extended region of the brightest infrared structure in the W4 band. This again suggests that the W3 and W4 emission is dominated by a W1 and W2 dropout radio galaxy - typical characteristics of hot DOGs \citep{2015ApJ...804...27A,2015ApJ...805...90T}.

\citet{2023ApJ...956...34W} has proposed that for a Dyson Sphere, energy could be radiated away from the structure as a strong, coherent radio signal, serving as a mechanism to dissipate the significant heat generated by the absorption and use of stellar energy. Such emission, if present, would offer an interesting observational technosignature. Our e-MERLIN and EVN observations reveal no detectable radio emission at the position of candidate G. The deeper EVN observations show no emission at a 3$\sigma$ level $ > 0.165 $~mJy/beam.

\section{Conclusion}
\label{sec5}

High-resolution e-MERLIN and EVN radio observations of J2335-0004, associated with Project Hephaistos Dyson Sphere candidate G, reveal compelling evidence that the mid-infrared excess previously attributed to a potential Dyson Sphere is instead the result of contamination by a background AGN. The combination of e-MERLIN and EVN data demonstrates that the source exhibits the typical characteristics of a radio-loud AGN in terms of its brightness temperature, component
spectral indices, and source morphology. 

Analysis of the positional offsets between the radio, mid-infrared (WISE), and optical ({\it{Gaia}}) data reinforces this interpretation. While the W1 and W2 infrared bands align closely with the M-dwarf star, the W3 and W4 bands are offset towards the AGN (J2335-0004), indicating that the latter dominates the mid-infrared emission. This contamination explains the spectral energy distribution anomaly that initially identified the G as a Dyson Sphere candidate.

Our findings align with the hypothesis that some, if not all, of the Dyson Sphere candidates from Project Hephaistos are affected by similar background AGN contamination. The likely culprits are "hot, dust-obscured galaxies" (hot DOGs), which are faint in the optical and NIR but are well detected in the mid-infrared. The absence of X-ray counterparts further supports the scenario of a heavily obscured, Compton thick AGN, also consistent with the properties of hot DOGs.

Finally, we note the absence of any detectable radio emission at the precise {\it{Gaia}} position of candidate G. All these findings underscore the critical role of high-resolution radio imaging in clarifying the origins of anomalous astrophysical data and highlight the necessity of meticulous background subtraction, particularly in the infrared, for the robust identification of Dyson Sphere candidates. Future high-resolution radio observations targeting other Project Hephaistos Dyson Sphere candidates ({\it e.g.}  candidates A and B), would be invaluable for further testing the selection criteria and distinguishing truly promising candidates from those affected by background contamination. The challenge of differentiating genuine mid-infrared excess sources from background AGN contamination is equally significant in searches for Extreme Debris Disks (EDDs) \citep{2021ApJ...910...27M,2024AJ....168..157C}. This highlights the need for careful source characterisation and follow-up observations to mitigate the risk of false positives in such studies.

\section*{Acknowledgements}

We thank Drs Benito Marcote, David Williams-Baldwin, Zsolt Paragi, Bob Campbell and, and Prof. Simon Garrington for their generous help with the e-MERLIN and EVN observations. We also thank Kelvin Wandia, Ramiro Saide and Dr Anthony Holloway for discussions on some aspects of the data analysis process. The European VLBI Network is a joint facility of European, Chinese, South African, and other radio astronomy institutes funded by their national research councils. e-MERLIN is a National Facility operated by the University of Manchester at Jodrell Bank Observatory on behalf of STFC. T.R. is funded by the China Scholarship Council.

%%%%%%%%%%%%%%%%%%%%%%%%%%%%%%%%%%%%%%%%%%%%%%%%%%
\section*{Data Availability}

This work used software: \textsc{Astropy}\citep{astropy:2013, astropy:2018, astropy:2022}, \textsc{NumPy} \citep{harris2020array}, Photutils \citep{larry_bradley_2024_12585239}, \textsc{Matplotlib} \citep{2007CSE.....9...90H}, \textsc{casa} \citep{2022PASP..134k4501C}, AIPS \citep{1985daa..conf..195W}. 

This work used data from {\it Gaia} \citep{2016A&A...595A...1G,2023A&A...674A...1G}, AllWISE Images Atlas \citep{2020ipac.data.I153W}, the Two Micron All Sky Survey (2MASS) \citep{2006AJ....131.1163S}, Aladin Sky Atlas\citep{2000A&AS..143...33B}, \textsc{Simbad}\citep{2000A&AS..143....9W}, HEASARC, Very Large Array Sky Survey (VLASS)\citep{2020PASP..132c5001L}. 

The data supporting this article will be made available by the corresponding author upon reasonable request.



\bibliographystyle{mnras}
\bibliography{example} % if your bibtex file is called example.bib

\begin{thebibliography}{}
\makeatletter
\relax
\def\mn@urlcharsother{\let\do\@makeother \do\$\do\&\do\#\do\^\do\_\do\%\do\~}
\def\mn@doi{\begingroup\mn@urlcharsother \@ifnextchar [ {\mn@doi@} {\mn@doi@[]}}
\def\mn@doi@[#1]#2{\def\@tempa{#1}\ifx\@tempa\@empty \href {http://dx.doi.org/#2} {doi:#2}\else \href {http://dx.doi.org/#2} {#1}\fi \endgroup}
\def\mn@eprint#1#2{\mn@eprint@#1:#2::\@nil}
\def\mn@eprint@arXiv#1{\href {http://arxiv.org/abs/#1} {{\tt arXiv:#1}}}
\def\mn@eprint@dblp#1{\href {http://dblp.uni-trier.de/rec/bibtex/#1.xml} {dblp:#1}}
\def\mn@eprint@#1:#2:#3:#4\@nil{\def\@tempa {#1}\def\@tempb {#2}\def\@tempc {#3}\ifx \@tempc \@empty \let \@tempc \@tempb \let \@tempb \@tempa \fi \ifx \@tempb \@empty \def\@tempb {arXiv}\fi \@ifundefined {mn@eprint@\@tempb}{\@tempb:\@tempc}{\expandafter \expandafter \csname mn@eprint@\@tempb\endcsname \expandafter{\@tempc}}}

\bibitem[\protect\citeauthoryear{{Assef} et~al.,}{{Assef} et~al.}{2015}]{2015ApJ...804...27A}
{Assef} R.~J.,  et~al., 2015, \mn@doi [\apj] {10.1088/0004-637X/804/1/27}, \href {https://ui.adsabs.harvard.edu/abs/2015ApJ...804...27A} {804, 27}

\bibitem[\protect\citeauthoryear{{Astropy Collaboration} et~al.,}{{Astropy Collaboration} et~al.}{2013}]{astropy:2013}
{Astropy Collaboration} et~al., 2013, \mn@doi [\aap] {10.1051/0004-6361/201322068}, \href {http://adsabs.harvard.edu/abs/2013A%26A...558A..33A} {558, A33}

\bibitem[\protect\citeauthoryear{{Astropy Collaboration} et~al.,}{{Astropy Collaboration} et~al.}{2018}]{astropy:2018}
{Astropy Collaboration} et~al., 2018, \mn@doi [\aj] {10.3847/1538-3881/aabc4f}, \href {https://ui.adsabs.harvard.edu/abs/2018AJ....156..123A} {156, 123}

\bibitem[\protect\citeauthoryear{{Astropy Collaboration} et~al.,}{{Astropy Collaboration} et~al.}{2022}]{astropy:2022}
{Astropy Collaboration} et~al., 2022, \mn@doi [\apj] {10.3847/1538-4357/ac7c74}, \href {https://ui.adsabs.harvard.edu/abs/2022ApJ...935..167A} {935, 167}

\bibitem[\protect\citeauthoryear{{Balog}, {Kiss}, {Vink{\'o}}, {Rieke}, {Muzerolle}, {G{\'a}sp{\'a}r}, {Young}  \& {Gorlova}}{{Balog} et~al.}{2009}]{2009ApJ...698.1989B}
{Balog} Z.,  {Kiss} L.~L.,  {Vink{\'o}} J.,  {Rieke} G.~H.,  {Muzerolle} J.,  {G{\'a}sp{\'a}r} A.,  {Young} E.~T.,   {Gorlova} N.,  2009, \mn@doi [\apj] {10.1088/0004-637X/698/2/1989}, \href {https://ui.adsabs.harvard.edu/abs/2009ApJ...698.1989B} {698, 1989}

\bibitem[\protect\citeauthoryear{{Blain}}{{Blain}}{2024}]{2024arXiv240911447B}
{Blain} A.~W.,  2024, \mn@doi [arXiv e-prints] {10.48550/arXiv.2409.11447}, \href {https://ui.adsabs.harvard.edu/abs/2024arXiv240911447B} {p. arXiv:2409.11447}

\bibitem[\protect\citeauthoryear{{Bonnarel} et~al.,}{{Bonnarel} et~al.}{2000}]{2000A&AS..143...33B}
{Bonnarel} F.,  et~al., 2000, \mn@doi [\aaps] {10.1051/aas:2000331}, \href {https://ui.adsabs.harvard.edu/abs/2000A&AS..143...33B} {143, 33}

\bibitem[\protect\citeauthoryear{Bradley et~al.,}{Bradley et~al.}{2024}]{larry_bradley_2024_12585239}
Bradley L.,  et~al., 2024, astropy/photutils: 1.13.0, \mn@doi{10.5281/zenodo.12585239}, \url {https://doi.org/10.5281/zenodo.12585239}

\bibitem[\protect\citeauthoryear{{CASA Team} et~al.,}{{CASA Team} et~al.}{2022}]{2022PASP..134k4501C}
{CASA Team} et~al., 2022, \mn@doi [\pasp] {10.1088/1538-3873/ac9642}, \href {https://ui.adsabs.harvard.edu/abs/2022PASP..134k4501C} {134, 114501}

\bibitem[\protect\citeauthoryear{{Condon}, {Huang}, {Yin}  \& {Thuan}}{{Condon} et~al.}{1991}]{1991ApJ...378...65C}
{Condon} J.~J.,  {Huang} Z.~P.,  {Yin} Q.~F.,   {Thuan} T.~X.,  1991, \mn@doi [\apj] {10.1086/170407}, \href {https://ui.adsabs.harvard.edu/abs/1991ApJ...378...65C} {378, 65}

\bibitem[\protect\citeauthoryear{{Contardo} \& {Hogg}}{{Contardo} \& {Hogg}}{2024}]{2024AJ....168..157C}
{Contardo} G.,  {Hogg} D.~W.,  2024, \mn@doi [\aj] {10.3847/1538-3881/ad6b90}, \href {https://ui.adsabs.harvard.edu/abs/2024AJ....168..157C} {168, 157}

\bibitem[\protect\citeauthoryear{{Cotten} \& {Song}}{{Cotten} \& {Song}}{2016}]{2016ApJS..225...15C}
{Cotten} T.~H.,  {Song} I.,  2016, \mn@doi [\apjs] {10.3847/0067-0049/225/1/15}, \href {https://ui.adsabs.harvard.edu/abs/2016ApJS..225...15C} {225, 15}

\bibitem[\protect\citeauthoryear{{Gaia Collaboration} et~al.,}{{Gaia Collaboration} et~al.}{2016}]{2016A&A...595A...1G}
{Gaia Collaboration} et~al., 2016, \mn@doi [\aap] {10.1051/0004-6361/201629272}, \href {https://ui.adsabs.harvard.edu/abs/2016A&A...595A...1G} {595, A1}

\bibitem[\protect\citeauthoryear{{Gaia Collaboration} et~al.,}{{Gaia Collaboration} et~al.}{2023}]{2023A&A...674A...1G}
{Gaia Collaboration} et~al., 2023, \mn@doi [\aap] {10.1051/0004-6361/202243940}, \href {https://ui.adsabs.harvard.edu/abs/2023A&A...674A...1G} {674, A1}

\bibitem[\protect\citeauthoryear{{Garrett} \& {Siemion}}{{Garrett} \& {Siemion}}{2023}]{2023MNRAS.519.4581G}
{Garrett} M.~A.,  {Siemion} A.~P.~V.,  2023, \mn@doi [\mnras] {10.1093/mnras/stac2607}, \href {https://ui.adsabs.harvard.edu/abs/2023MNRAS.519.4581G} {519, 4581}

\bibitem[\protect\citeauthoryear{Harris et~al.,}{Harris et~al.}{2020}]{harris2020array}
Harris C.~R.,  et~al., 2020, \mn@doi [Nature] {10.1038/s41586-020-2649-2}, 585, 357

\bibitem[\protect\citeauthoryear{{Hunter}}{{Hunter}}{2007}]{2007CSE.....9...90H}
{Hunter} J.~D.,  2007, \mn@doi [Computing in Science and Engineering] {10.1109/MCSE.2007.55}, \href {https://ui.adsabs.harvard.edu/abs/2007CSE.....9...90H} {9, 90}

\bibitem[\protect\citeauthoryear{{Lacy} et~al.,}{{Lacy} et~al.}{2020}]{2020PASP..132c5001L}
{Lacy} M.,  et~al., 2020, \mn@doi [\pasp] {10.1088/1538-3873/ab63eb}, \href {https://ui.adsabs.harvard.edu/abs/2020PASP..132c5001L} {132, 035001}

\bibitem[\protect\citeauthoryear{{Miville-Desch{\^e}nes} \& {Lagache}}{{Miville-Desch{\^e}nes} \& {Lagache}}{2005}]{2005ApJS..157..302M}
{Miville-Desch{\^e}nes} M.-A.,  {Lagache} G.,  2005, \mn@doi [\apjs] {10.1086/427938}, \href {https://ui.adsabs.harvard.edu/abs/2005ApJS..157..302M} {157, 302}

\bibitem[\protect\citeauthoryear{{Moldon}}{{Moldon}}{2021}]{2021ascl.soft09006M}
{Moldon} J.,  2021, {eMCP: e-MERLIN CASA pipeline}, Astrophysics Source Code Library, record ascl:2109.006

\bibitem[\protect\citeauthoryear{{Mo{\'o}r} et~al.,}{{Mo{\'o}r} et~al.}{2021}]{2021ApJ...910...27M}
{Mo{\'o}r} A.,  et~al., 2021, \mn@doi [\apj] {10.3847/1538-4357/abdc26}, \href {https://ui.adsabs.harvard.edu/abs/2021ApJ...910...27M} {910, 27}

\bibitem[\protect\citeauthoryear{{Ren}, {Garrett}  \& {Siemion}}{{Ren} et~al.}{2024}]{2024RNAAS...8..145R}
{Ren} T.,  {Garrett} M.~A.,   {Siemion} A. P.~V.,  2024, \mn@doi [Research Notes of the American Astronomical Society] {10.3847/2515-5172/ad5017}, \href {https://ui.adsabs.harvard.edu/abs/2024RNAAS...8..145R} {8, 145}

\bibitem[\protect\citeauthoryear{{Reynolds}, {Paragi}  \& {Garrett}}{{Reynolds} et~al.}{2002}]{2002astro.ph..5118R}
{Reynolds} C.,  {Paragi} Z.,   {Garrett} M.,  2002, \mn@doi [arXiv e-prints] {10.48550/arXiv.astro-ph/0205118}, \href {https://ui.adsabs.harvard.edu/abs/2002astro.ph..5118R} {pp astro--ph/0205118}

\bibitem[\protect\citeauthoryear{{Skrutskie} et~al.,}{{Skrutskie} et~al.}{2006}]{2006AJ....131.1163S}
{Skrutskie} M.~F.,  et~al., 2006, \mn@doi [\aj] {10.1086/498708}, \href {https://ui.adsabs.harvard.edu/abs/2006AJ....131.1163S} {131, 1163}

\bibitem[\protect\citeauthoryear{{Stern} et~al.,}{{Stern} et~al.}{2014}]{2014ApJ...794..102S}
{Stern} D.,  et~al., 2014, \mn@doi [\apj] {10.1088/0004-637X/794/2/102}, \href {https://ui.adsabs.harvard.edu/abs/2014ApJ...794..102S} {794, 102}

\bibitem[\protect\citeauthoryear{{Suazo}, {Zackrisson}, {Mahto}, {Lundell}, {Nettelblad}, {Korn}, {Wright}  \& {Majumdar}}{{Suazo} et~al.}{2024}]{2024MNRAS.531..695S}
{Suazo} M.,  {Zackrisson} E.,  {Mahto} P.~K.,  {Lundell} F.,  {Nettelblad} C.,  {Korn} A.~J.,  {Wright} J.~T.,   {Majumdar} S.,  2024, \mn@doi [\mnras] {10.1093/mnras/stae1186}, \href {https://ui.adsabs.harvard.edu/abs/2024MNRAS.531..695S} {531, 695}

\bibitem[\protect\citeauthoryear{{Tsai} et~al.,}{{Tsai} et~al.}{2015}]{2015ApJ...805...90T}
{Tsai} C.-W.,  et~al., 2015, \mn@doi [\apj] {10.1088/0004-637X/805/2/90}, \href {https://ui.adsabs.harvard.edu/abs/2015ApJ...805...90T} {805, 90}

\bibitem[\protect\citeauthoryear{{WISE Team}}{{WISE Team}}{2020}]{2020ipac.data.I153W}
{WISE Team} 2020, {AllWISE Atlas (L3a) Coadd Images}, NASA IPAC DataSet, IRSA153, \mn@doi{10.26131/IRSA153}

\bibitem[\protect\citeauthoryear{{Wells}}{{Wells}}{1985}]{1985daa..conf..195W}
{Wells} D.~C.,  1985, in {di Gesu} V.,  {Scarsi} L.,  {Crane} P.,  {Friedman} J.~H.,   {Levialdi} S.,  eds, Data Analysis in Astronomy. p.~195

\bibitem[\protect\citeauthoryear{{Wenger} et~al.,}{{Wenger} et~al.}{2000}]{2000A&AS..143....9W}
{Wenger} M.,  et~al., 2000, \mn@doi [\aaps] {10.1051/aas:2000332}, \href {https://ui.adsabs.harvard.edu/abs/2000A&AS..143....9W} {143, 9}

\bibitem[\protect\citeauthoryear{{Wlodarczyk-Sroka}, {Garrett}  \& {Siemion}}{{Wlodarczyk-Sroka} et~al.}{2020}]{2020MNRAS.498.5720W}
{Wlodarczyk-Sroka} B.~S.,  {Garrett} M.~A.,   {Siemion} A.~P.~V.,  2020, \mn@doi [\mnras] {10.1093/mnras/staa2672}, \href {https://ui.adsabs.harvard.edu/abs/2020MNRAS.498.5720W} {498, 5720}

\bibitem[\protect\citeauthoryear{{Wright}}{{Wright}}{2023}]{2023ApJ...956...34W}
{Wright} J.~T.,  2023, \mn@doi [\apj] {10.3847/1538-4357/acf44f}, \href {https://ui.adsabs.harvard.edu/abs/2023ApJ...956...34W} {956, 34}

\makeatother
\end{thebibliography}

%\begin{thebibliography}{99}
%\end{thebibliography}

%%%%%%%%%%%%%%%%%%%%%%%%%%%%%%%%%%%%%%%%%%%%%%%%%%

%%%%%%%%%%%%%%%%% APPENDICES %%%%%%%%%%%%%%%%%%%%%

%\appendix

%%%%%%%%%%%%%%%%%%%%%%%%%%%%%%%%%%%%%%%%%%%%%%%%%%

% Don't change these lines
\bsp	% typesetting comment
\label{lastpage}
\end{document}